\documentclass[aps,preprint]{revtex4}%
\usepackage{amsfonts}
\usepackage{amsmath}
\usepackage{amssymb}
\usepackage{graphicx}%
\newtheorem{theorem}{Theorem}

\newtheorem{remark}[theorem]{Remark}

\begin{document}
\preprint{ }
\title{Theorem to generate Einstein-Non Linear Maxwell Fields}
\author{S. Habib Mazharimousavi}
\email{habib.mazhari@emu.edu.tr}
\author{O. Gurtug}
\email{ozay.gurtug@emu.edu.tr}
\author{M. Halilsoy}
\email{mustafa.halilsoy@emu.edu.tr}
\affiliation{Department of Physics, Eastern Mediterranean University, G. Magusa, north
Cyprus, Mersin 10 - Turkey.}
\pacs{PACS number}

\begin{abstract}
We present a theorem in d-dimensional static, spherically symmetric spacetime
in generic Lovelock gravity coupled with a non-linear electrodynamic source to
generate solutions. The theorem states that irrespective of the order \ of
\ the Lovelock gravity and non-linear Maxwell (NLM) Lagrangian, for the pure
electric field case the NLM equations are satisfied by virtue of the
Einstein-Lovelock equations. Applications of the theorem, specifically to the
study of black hole solutions in Chern-Simons (CS) theory is given. Radiating
version of the theorem has been considered, which generalizes the
Bonnor-Vaidya (BV) metric to the Lovelock gravity with a NLM field as a
radiating source. We consider also the radiating power - Maxwell source ( i.e.
$\left(  F_{\mu\nu}F^{\mu\nu}\right)  ^{q},$ $q=$ finely - tuned constant )
within the context of Lovelock gravity.

\end{abstract}
\maketitle

\section{Introduction}

The string theory motivated higher dimensional gravity, known as the Lovelock
gravity \cite{1} attracted much interest in recent years \cite{2}. This theory
is known to admit the most general higher order invariants in such
combinations that the field equations preserve their second order form. These
are important features concerning divergences at smaller scales and ghost free
structure toward a quantum theory of gravity. The first few orders of the
Lovelock Lagrangian are well known: zeroth /first order is just the
cosmological constant ($\Lambda$) /Einstein-Hilbert (EH) term. The second
order Lovelock term is known also as the Gauss-Bonnet (GB) term, which
consists of quadratic invariants and is a highly non-trivial contribution,
especially in higher dimensions\cite{3}. Going to higher order corrections
serves only to add contributions at higher levels of intricacy which are
crucial in a possible quantum gravity. In spite of all complications it is
remarkable that in a spherically symmetric line element exact black hole
solutions are found in higher order Lovelock gravity \cite{4}. These metrics
are sourced by Maxwell and Yang-Mills (YM) fields. More recently, we obtained
black holes in a power-YM source in Lovelock theory\cite{5}. By the power-YM
source it is meant that the source consists of a power of the YM invariant,
i.e., $\left(  F_{\mu\nu}^{a}F^{a\mu\nu}\right)  ^{q}$. Here $F_{\mu\nu}^{a}$
refers to the YM field with gauge group index a and $q>0$ is an arbitrary
constant parameter. Although $q$ may stand arbitrary, energy and causality
conditions restrict it to a certain set of admissible parameters \cite{5}.

By the same token, in this paper we consider Lovelock gravity of all higher
orders coupled with a non-linear Maxwell (NLM) source\cite{6}. We prove a
Theorem, covering Lovelock terms to all orders - albeit with proper constant
coefficients - which generates new metrics once the NLM energy-momentum is
known. The energy-momentum is given by $\mathcal{L}\left(  \mathcal{F}\right)
,$ in which $\mathcal{F}=F_{\mu\nu}F^{\mu\nu}$ denotes the Maxwell invariant.
The field equation satisfied by $F_{\mu\nu}$ will be referred to as the NLM
equation. The Theorem involves spherically symmetric metric ansatz together
with the NLM Lagrangian which leads to a general class of metrics. Static
electric fields fall within the range of our Theorem which yields Born-Infeld
(BI) electrodynamics as a particular example. A static, pure magnetic YM field
constitutes another example of BI type.

For the first/second order Lovelock terms the Theorem addresses to EH/GB
gravities, resulting in metrics expressed in solutions of first/second order
algebraic equations. With equal ease, we extend this result to any higher
order Lovelock gravity in terms of a higher order algebraic equations
\cite{7}. As an interesting example we obtain Chern-Simons-Born-Infeld (CSBI)
black hole solutions in odd dimensions. We elaborate on $d=5$, to show in
particular that the thermodynamically well-behaving CS black hole preserves,
its good features when coupled with a BI source in the general relativity
limit. We note that combination of CS and BI types is in an artificial manner
since CS/BI black holes arise naturally in odd /even dimensions. Such a
combination works only when the $^{\star}F_{\mu\nu}$ (i.e., dual of $F_{\mu
\nu}$) vanishes, which occurs in a restricted type of sources. BI black holes
are known to arise naturally from Pfaffians only in even dimensional
spacetimes \cite{5}. Depending on the topological parameter $\chi=0,\pm1$ and
$\Lambda\gtrless0$ we investigate the Hawking temperature of the 5-dimensional
CSBI black holes. Finally, we investigate the implications of our Theorem in
the Eddington-Bondi form of radiating metrics \cite{8}. This is the time
dependent version of both the mass and charge so that the metrics become time
dependent. That is, we extend the Bonnor-Vaidya \cite{8} form of radiating
metric to the d-dimensional Lovelock gravity coupled with a NLM field. We
consider various types of non-linearities in electromagnetic field as
particular examples.

Organization of the paper is as follows. The Theorem, its proof and
applications are given in Sec. II. Generalization of the Theorem to the
radiating metrics in higher dimensions with NLM as source is studied in Sec.
III. In section IV, we investigate the case of power-Maxwell non-linearity.
The paper is completed with the conclusion in Sec. V.

\section{A theorem for solving Einstein-Lovelock-NLM equations}

\textbf{Theorem :} \emph{Let the d-dimensional static spherically symmetric
spacetime sourced by non-linear electromagnetic field be described by the
action }$\left(  8\pi G=1\right)  $%
\begin{equation}
S=\frac{1}{2}\int dx^{d}\sqrt{-g}\left\{  -\frac{\left(  d-2\right)  \left(
d-1\right)  }{3}\Lambda+%
\mathcal{L}%
_{1}+\alpha_{2}%
\mathcal{L}%
_{2}+...+\alpha_{\left[  \frac{d-1}{2}\right]  }%
\mathcal{L}%
_{\left[  \frac{d-1}{2}\right]  }+\mathcal{L}\left(  \mathcal{F}\right)
\right\}  \ \ ,
\end{equation}
\emph{in which }%
\begin{equation}%
\mathcal{L}%
_{n}=2^{-n}\delta_{c_{1}...c_{n}d_{1}...d_{n}}^{a_{1}...a_{n}b_{1}...b_{n}%
}R_{\ \ a_{1}b_{1}}^{c_{1}d_{1}}...R_{\ \ a_{n}b_{n}}^{c_{n}d_{n}}%
,\ \ \ n\geq1,
\end{equation}
\emph{is the }$n$\emph{th order Lovelock Lagrangian, }$\alpha_{n}$\emph{\ is a
real constant and the bracket }$[.]$\emph{\ refers to integer part}.
\emph{Herein }$\mathcal{F}=F_{\mu\nu}F^{\mu\nu},$\emph{\ is the Maxwell
invariant for a static field 2-form }%
\begin{equation}
F=E\left(  r\right)  dt\wedge dr.
\end{equation}
\emph{If the }$\mathcal{L}\left(  \mathcal{F}\right)  $\emph{\ satisfies the
non-linear Maxwell (NLM) equation}%
\begin{equation}
d\left(  ^{\star}F\mathcal{L}_{\mathcal{F}}\right)  =0
\end{equation}
\emph{with }$\mathcal{L}_{\mathcal{F}}=\frac{\partial\mathcal{L}}%
{\partial\mathcal{F}}$ \emph{and }$^{\star}F$ \emph{the dual of }$F$,
\emph{then the Einstein equations admit the solution}%
\begin{equation}
ds^{2}=-(\chi-r^{2}H\left(  r\right)  )dt^{2}+\frac{1}{(\chi-r^{2}H\left(
r\right)  )}dr^{2}+r^{2}d\Omega_{d-2}^{2}%
\end{equation}
\emph{where }$\chi=0,\pm1$ \emph{and in which} $H\left(  r\right)  $ \emph{is
the solution (or solutions) of the following algebraic equation of order}
$\left[  \frac{d-1}{2}\right]  $
\begin{equation}
\sum_{k=1}^{\left[  \frac{d-1}{2}\right]  }\tilde{\alpha}_{k}H\left(
r\right)  ^{k}=\frac{\Lambda}{3}+\frac{M}{r^{d-1}}-\frac{1}{\left(
d-2\right)  r^{d-1}}%
{\textstyle\int}
r^{d-2}\left(  \mathcal{L}-2\mathcal{L}_{\mathcal{F}}\mathcal{F}\right)  dr.
\end{equation}
\emph{Here} $m$\emph{\ is an integration constant, }$\tilde{\alpha}_{k}=%
{\textstyle\prod\limits_{i=3}^{2k}}
\left(  d-i\right)  \alpha_{k}$ and $\tilde{\alpha}_{1}=1.$\emph{\ }

\textbf{Proof}: Variation of the action with respect to the metric tensor
$g_{\mu\nu}$ yields the field equations in the form%

\begin{equation}
\mathcal{G}_{\mu}^{\ \nu}\left(  =\mathcal{G}_{\mu}^{\ \nu(EH)}+\sum
_{k=2}^{\left[  \frac{d-1}{2}\right]  }\alpha_{k}\mathcal{G}_{\mu}%
^{\ \nu\left(  k\right)  }\right)  +\frac{\left(  d-2\right)  \left(
d-1\right)  }{6}\Lambda\delta_{\mu}^{\ \nu}=T_{\mu}^{\ \nu},
\end{equation}
where $\mathcal{G}_{\mu}^{\ \nu(EH)}$ is the Einstein tensor, while%

\begin{equation}
\mathcal{G}_{\mu}^{\ \nu\left(  k\right)  }=\frac{1}{2^{k+1}}\delta
_{bc_{1}...c_{k}d_{1}...d_{k}}^{aa_{1}...a_{k}b_{1}...b_{k}}R_{\ \ a_{1}b_{1}%
}^{c_{1}d_{1}}...R_{\ \ a_{k}b_{k}}^{c_{k}d_{k}}.
\end{equation}
The $T_{\mu}^{\ \nu}$ is given by
\begin{equation}
T_{\mu}^{\ \nu}=\frac{1}{2}\left(  \mathcal{L}\delta_{\mu}^{\ \nu
}-4\mathcal{L}_{\mathcal{F}}F_{\mu\lambda}F^{\nu\lambda}\right)  ,
\end{equation}
which clearly gives $T_{t}^{\ t}=T_{r}^{\ r}=\frac{1}{2}\mathcal{L}%
-\mathcal{L}_{\mathcal{F}}\mathcal{F},$ stating that $\mathcal{G}_{t}%
^{\ t}=\mathcal{G}_{r}^{\ r}$ and $T_{\theta_{i}}^{\ \theta_{i}}=\frac{1}%
{2}\mathcal{L}$. Now we introduce our metric as given by (5), where the choice
of $g_{tt}=-\left(  g_{rr}\right)  ^{-1}$ is a direct result of $\mathcal{G}%
_{t}^{\ t}=\mathcal{G}_{r}^{\ r}$ up to a constant coefficient where we set it
to be one. By starting with the line element (5) one gets%
\begin{equation}
\mathcal{G}_{t}^{t}=\mathcal{G}_{r}^{r}=-\frac{\left(  d-2\right)  }{2r^{d-2}%
}\left(  r^{d-1}H\left(  r\right)  \right)  ^{\prime}-\frac{\left(
d-2\right)  }{2r^{d-2}}\sum_{k=2}^{\left[  \frac{d-1}{2}\right]  }%
\tilde{\alpha}_{k}\left(  r^{d-1}H\left(  r\right)  ^{k}\right)  ^{\prime},
\end{equation}
and
\begin{equation}
\mathcal{G}_{\theta_{i}}^{\ \theta_{i}}=-\frac{1}{2r^{d-3}}\left(
r^{d-1}H\left(  r\right)  \right)  ^{\prime\prime}-\frac{1}{2r^{d-3}}%
\sum_{k=2}^{\left[  \frac{d-1}{2}\right]  }\tilde{\alpha}_{k}\left(
r^{d-1}H\left(  r\right)  ^{k}\right)  ^{\prime\prime},
\end{equation}
where $\left(  .\right)  ^{\prime}=\frac{d}{dr}\left(  .\right)  .$ Eq.s (10)
and (11) admit
\begin{equation}
\left(  \frac{r^{d-2}}{d-2}\mathcal{G}_{t}^{\ t}\right)  ^{\prime}%
=r^{d-3}\mathcal{G}_{\theta_{i}}^{\ \theta_{i}},
\end{equation}
and imposing Einstein equations yield
\begin{equation}
\left(  \frac{r^{d-2}}{d-2}\left(  T_{t}^{\ t}-\frac{\left(  d-2\right)
\left(  d-1\right)  }{6}\Lambda\right)  \right)  ^{\prime}=r^{d-3}\left(
T_{\theta_{i}}^{\ \theta_{i}}-\frac{\left(  d-2\right)  \left(  d-1\right)
}{6}\Lambda\right)  .
\end{equation}
This is equivalent to,
\begin{equation}
\left(  \frac{r^{d-2}}{d-2}\left(  \frac{1}{2}\mathcal{L}-\mathcal{L}%
_{\mathcal{F}}\mathcal{F}\right)  \right)  ^{\prime}=r^{d-3}\left(  \frac
{1}{2}\mathcal{L}\right)  ,
\end{equation}
which integrates to
\begin{equation}
r^{d-2}\mathcal{L}_{\mathcal{F}}\sqrt{\left\vert \mathcal{F}\right\vert
}=\text{constant.}%
\end{equation}
This \ result is in conform with the solution of the NLM equation (4) and the
integration constant can be identified with the electric charge. Now recall
from (7), that
\begin{equation}
\mathcal{G}_{t}^{\ t}+\frac{\left(  d-2\right)  \left(  d-1\right)  }%
{6}\Lambda=T_{t}^{\ t},
\end{equation}
or equivalently
\begin{equation}
-\frac{\left(  d-2\right)  }{2r^{d-2}}\left(  r^{d-1}H\left(  r\right)
\right)  ^{\prime}-\frac{\left(  d-2\right)  }{2r^{d-2}}\sum_{k=2}^{\left[
\frac{d-1}{2}\right]  }\tilde{\alpha}_{k}\left(  r^{d-1}H\left(  r\right)
^{k}\right)  ^{\prime}=\frac{1}{2}\mathcal{L}-\mathcal{L}_{\mathcal{F}%
}\mathcal{F}-\frac{\left(  d-2\right)  \left(  d-1\right)  }{6}\Lambda,
\end{equation}
which implies
\begin{equation}
\sum_{k=1}^{\left[  \frac{d-1}{2}\right]  }\tilde{\alpha}_{k}H\left(
r\right)  ^{k}=\frac{\Lambda}{3}+\frac{M}{r^{d-1}}-\frac{1}{\left(
d-2\right)  r^{d-1}}%
{\textstyle\int}
r^{d-2}\left(  \mathcal{L}-2\mathcal{L}_{\mathcal{F}}\mathcal{F}\right)  dr,
\end{equation}
for the integration constant $M=\frac{4m}{\left(  d-2\right)  }.$ This
completes the proof of our Theorem.

\begin{remark}
In the Theorem, we considered only a static electric field whose 2-form is
$F=E\left(  r\right)  dt\wedge dr,$ giving $\mathcal{F=}-2E^{2}.$ Since
$\mathcal{F}$ is only a function of $r$ so is the Lagrangian $\mathcal{L}%
\left(  \mathcal{F}\right)  ,$ this aided in the proof of the Theorem. We must
add that the Theorem becomes of practical use whenever the expression on the
right hand side is integrable. Otherwise we should resort to the multipole
expansions for the asymptotically flat spacetimes.
\end{remark}

\begin{remark}
In the case of pure magnetic field \ the Theorem is applicable only if
$\mathcal{F}=F_{\mu\nu}F^{\mu\nu}$ is only a function of $r.$ This implies
\begin{equation}
\left(  \frac{r^{d-2}}{d-2}T_{t}^{\ t}\right)  ^{\prime}=r^{d-3}T_{\theta_{i}%
}^{\ \theta_{i}},
\end{equation}
which leads to the same metric function in the form
\begin{equation}
\sum_{k=1}^{\left[  \frac{d-1}{2}\right]  }\tilde{\alpha}_{k}H\left(
r\right)  ^{k}=\frac{\Lambda}{3}+\frac{M}{r^{d-1}}-\frac{2}{\left(
d-2\right)  r^{d-1}}%
{\textstyle\int}
r^{d-2}T_{t}^{\ t}dr.
\end{equation}

\end{remark}

\begin{remark}
In the case of a general energy - momentum tensor
\begin{equation}
T_{\mu}^{\nu}=diag\left(  T_{t}^{\ t},T_{r}^{\ r},T_{\theta_{1}}^{\ \theta
_{1}},...\right)  ,
\end{equation}
in which
\begin{equation}
T_{t}^{\ t}=T_{r}^{\ r},T_{\theta_{1}}^{\ \theta_{1}}=T_{\theta_{2}}%
^{\ \theta_{2}}=...
\end{equation}
and
\begin{equation}
\left(  \frac{r^{d-2}}{d-2}T_{t}^{\ t}\right)  ^{\prime}=r^{d-3}T_{\theta_{i}%
}^{\ \theta_{i}},
\end{equation}
again, a solution in the form of (5) satisfies the Einstein equations with
$H\left(  r\right)  $ given by (20).
\end{remark}

\begin{remark}
The case of $\mathcal{L}-2\mathcal{L}_{\mathcal{F}}\mathcal{F}=0$ must be
excluded, since it implies a Lagrangian of the form $\mathcal{L=}%
\sqrt{\mathcal{F}},$ which fails to satisfy the energy and causality
conditions \cite{5}. This form of the Lagrangian lacks also the linear Maxwell limit.
\end{remark}

\textit{Example 1:} As an application we consider the case of pure electric
Einstein-Born-Infeld (EBI) black hole solution. The pure electric BI
Lagrangian can be written as \cite{6}
\begin{equation}
\mathcal{L}\left(  \mathcal{F}\right)  =4\beta^{2}\left(  1-\sqrt
{1+\frac{\mathcal{F}}{2\beta^{2}}}\right)  ,
\end{equation}
where $\mathcal{F}$ is the electric field invariant given by $\mathcal{F}%
=2F_{tr}F^{tr}=-2E\left(  r\right)  ^{2}.$ Note that for $\beta\rightarrow
\infty$ we recover the standard, linear Maxwell Lagrangian. Consequently
\begin{equation}
\mathcal{L}_{\mathcal{F}}=-\frac{1}{\sqrt{1+\frac{\mathcal{F}}{2\beta^{2}}}},
\end{equation}
and therefore upon solving the non-linear Maxwell equation one finds%
\begin{equation}
E=\frac{q\beta}{\sqrt{q^{2}+\beta^{2}r^{2\left(  d-2\right)  }}}.
\end{equation}
Finally we obtain
\begin{align}
H\left(  r\right)   &  =\frac{\Lambda}{3}+\frac{M}{r^{d-1}}-\frac{2}{\left(
d-2\right)  r^{d-1}}%
{\textstyle\int}
r^{d-2}T_{t}^{\ t}dr==\frac{\Lambda}{3}+\frac{4m}{\left(  d-2\right)  r^{d-1}%
}-\\
&  \frac{4\beta^{2}}{\left(  d-1\right)  \left(  d-2\right)  }\left(
1-\sqrt{1+\frac{q^{2}}{\beta^{2}r^{2\left(  d-2\right)  }}}\right)  -\frac
{4}{\left(  d-1\right)  \left(  d-3\right)  }\frac{q^{2}}{r^{2\left(
d-2\right)  }}\times\nonumber\\
&  \text{ }_{2}F_{1}\left(  \frac{1}{2},\frac{d-3}{2\left(  d-2\right)
},\frac{3d-7}{2\left(  d-2\right)  },-\frac{q^{2}}{\beta^{2}r^{2\left(
d-2\right)  }}\right)  ,\nonumber
\end{align}
in which $_{2}F_{1}$ stands for the hypergeometric function.

\textit{Example 2: }Another example for the case of non-electric field is
given by the Einstein--Yang-Mills (EYM) non-linear electrodynamics black hole
solution. In fact in Ref. \cite{9} we find that
\begin{equation}
T_{t}^{\ t}=T_{r}^{\ r}=2\beta^{2}\left(  1-\sqrt{1+\frac{\left(  d-2\right)
\left(  d-3\right)  Q^{2}}{2\beta^{2}r^{4}}}\right)  ,
\end{equation}
and
\begin{equation}
T_{\theta_{i}}^{\ \theta_{i}}=2\beta^{2}\left(  1-\sqrt{1+\frac{\left(
d-2\right)  \left(  d-3\right)  Q^{2}}{2\beta^{2}r^{4}}}\right)
+\frac{2\left(  d-3\right)  Q^{2}}{r^{4}\sqrt{1+\frac{\left(  d-2\right)
\left(  d-3\right)  Q^{2}}{2\beta^{2}r^{4}}}},
\end{equation}
which clearly satisfies the conditions of the Theorem and the Einstein
equations admit a black hole solution with
\begin{align}
H\left(  r\right)   &  =\frac{\Lambda}{3}+\frac{4m}{\left(  d-2\right)
r^{d-1}}-\frac{4\beta^{2}}{\left(  d-1\right)  \left(  d-2\right)
}+\nonumber\\
&  \frac{4\beta^{2}}{\left(  d-2\right)  r^{d-1}}%
{\textstyle\int}
drr^{d-4}\sqrt{r^{4}+\frac{\left(  d-2\right)  \left(  d-3\right)  Q^{2}%
}{2\beta^{2}}},
\end{align}
in conform with the solution given in Ref. \cite{9}.

\textit{Example 3: }Our next example will be the general form of energy
momentum tensor given by Salgado \cite{10} which states that
\begin{equation}
T_{\mu\left(  Diag.\right)  }^{\nu}=\frac{C}{r^{n\left(  1-k\right)  }}\left[
1,1,k,...,k\right]  ,\text{ \ \ \ }(C,k:\text{constants}),\nonumber
\end{equation}
admits solutions for Einstein equations. Now we show that this is a natural
result for the case of Rem. 3. In Rem. 3 let's consider
\begin{equation}
T_{t}^{\ t}=T_{r}^{\ r},\text{ \ \ \ \ }T_{\theta_{i}}^{\ \theta_{i}}%
=kT_{t}^{\ t},
\end{equation}
then, from equation (23) we obtain%
\begin{equation}
\left(  \frac{r^{d-2}}{d-2}T_{t}^{\ t}\right)  ^{\prime}=kr^{d-3}T_{t}^{\ t},
\end{equation}
or, in a straightforward calculation one finds%
\begin{equation}
T_{t}^{\ t}=\frac{C}{r^{n\left(  1-k\right)  }},
\end{equation}
where $C$ is an integration constant. This verifies that the Theorem proved by
Salgado \cite{10} turns out to be a particular case of our more general Theorem.

\textit{Example 4}: One may notice that a proper choice of $\tilde{\alpha}%
_{k}$ leads to a Chern-Simons (CS) \cite{11} gravity in odd dimensions. To do
so we set
\begin{equation}
\tilde{\alpha}_{k}=\frac{\bar{\alpha}_{k}}{\bar{\alpha}_{1}},\text{ for }%
k\geq2\text{ and }-\frac{\Lambda}{3}=\frac{\bar{\alpha}_{0}}{\bar{\alpha}_{1}%
},
\end{equation}
and we rewrite (18) as%
\begin{equation}
\sum_{k=0}^{\left[  \frac{d-1}{2}\right]  }\bar{\alpha}_{k}H\left(  r\right)
^{k}=\bar{\alpha}_{1}\left(  \frac{M}{r^{d-1}}-\frac{1}{\left(  d-2\right)
r^{d-1}}%
{\textstyle\int}
r^{d-2}\left(  \mathcal{L}-2\mathcal{L}_{\mathcal{F}}\mathcal{F}\right)
dr\right)  .
\end{equation}
Now we choose
\begin{equation}
\bar{\alpha}_{k}=\binom{\left[  \frac{d-1}{2}\right]  }{k}\ell^{2k-d},
\end{equation}
where
\begin{equation}
-\frac{\Lambda}{3}=\frac{\bar{\alpha}_{0}}{\bar{\alpha}_{1}}=\frac{\ell^{-2}%
}{\left[  \frac{d-1}{2}\right]  },
\end{equation}
to get from the binomial expansion
\begin{equation}
\left(  1+\ell^{2}H\left(  r\right)  \right)  ^{\left[  \frac{d-1}{2}\right]
}=\ell^{d}\bar{\alpha}_{1}\left(  \frac{M}{r^{d-1}}-\frac{1}{\left(
d-2\right)  r^{d-1}}%
{\textstyle\int}
r^{d-2}\left(  \mathcal{L}-2\mathcal{L}_{\mathcal{F}}\mathcal{F}\right)
dr\right)  .
\end{equation}
This implies that
\begin{equation}
H\left(  r\right)  =-\frac{1}{\ell^{2}}+\frac{\sigma}{\ell^{2}}\left[
\ell^{d}\bar{\alpha}_{1}\left(  \frac{M}{r^{d-1}}-\frac{1}{\left(  d-2\right)
r^{d-1}}%
{\textstyle\int}
r^{d-2}\left(  \mathcal{L}-2\mathcal{L}_{\mathcal{F}}\mathcal{F}\right)
dr\right)  \right]  ^{\frac{1}{\left[  \frac{d-1}{2}\right]  }},
\end{equation}
where $\sigma=+1$ if $\left[  \frac{d-1}{2}\right]  $ is an odd integer and
$\sigma=\pm1$ if $\left[  \frac{d-1}{2}\right]  $ is an even integer.

The latter equation for $d=$odd, $\chi=1$ admits ($\ell^{2}>0$)%
\begin{equation}
f\left(  r\right)  =1+\frac{r^{2}}{\ell^{2}}-\sigma\left[  \ell\bar{\alpha
}_{1}\left(  M-\frac{2B\left(  r\right)  }{\left(  d-2\right)  }\right)
\right]  ^{\frac{2}{d-1}}%
\end{equation}
where $f\left(  r\right)  =(\chi-r^{2}H\left(  r\right)  )$ is the metric
function and $B\left(  r\right)  =%
{\textstyle\int^{r}}
z^{d-2}T_{t}^{t}\left(  z\right)  dz.$ This metric function by using (39) and
(40) becomes%
\begin{equation}
f\left(  r\right)  =1+\frac{r^{2}}{\ell^{2}}-\sigma\left[  m+1-\frac{\left(
d-1\right)  B\left(  r\right)  }{\left(  d-2\right)  \ell^{\left(  d-3\right)
}}\right]  ^{\frac{2}{d-1}}.
\end{equation}
in which the new integration constant $m$ is related to $M$ \cite{11}. In the
sequel we investigate some thermodynamic properties of this solution.

\subsubsection{With $\sigma=+1$ or $\frac{d-1}{2}$ is an odd integer
($d=7,11,15,...$)}

In this case one finds the Hawking's temperature as%
\begin{equation}
T_{H}=\frac{1}{4\pi}f^{\prime}\left(  r_{+}\right)  =\frac{1}{2\pi}\left(
\frac{r_{+}}{\ell^{2}}+\frac{r_{+}^{3}T_{t}^{t}\left(  r_{+}\right)  }{\left(
d-2\right)  \left(  \ell^{2}+r_{+}^{2}\right)  ^{\left(  \frac{d-3}{2}\right)
}}\right)  .
\end{equation}
The specific heat capacity of the black hole for constant charge is defined
by
\begin{equation}
C_{q}=T_{H}\left(  \frac{\partial S}{\partial T_{H}}\right)  _{q},\text{
\ \ \ \ \ \ \ \ }S=\frac{\left(  d-1\right)  \pi^{\frac{d-1}{2}}}{4\left(
\frac{d-1}{2}\right)  !}r_{+}^{d-2},
\end{equation}
and is obtained as
\begin{equation}
C_{q}=\frac{\left(  d-1\right)  \left(  d-2\right)  \pi^{\left(  \frac{d-1}%
{2}\right)  }r_{+}^{d-2}}{\Gamma\left(  \frac{d+1}{2}\right)  }\frac{\Upsilon
}{\Psi},
\end{equation}
where
\begin{align}
\Upsilon &  =\left(  d-2\right)  +\frac{r_{+}^{2}\ell^{2}}{\left(  \ell
^{2}+r_{+}^{2}\right)  ^{\left(  \frac{d-3}{2}\right)  }}T_{t}^{t}\left(
r_{+}\right)  ,\\
\Psi=\left(  d-2\right)  +\frac{\ell^{2}r_{+}^{2}}{\left(  \ell^{2}+r_{+}%
^{2}\right)  ^{\left(  \frac{d-1}{2}\right)  }}  &  \left[  r_{+}\left(
\ell^{2}+r_{+}^{2}\right)  \frac{\partial}{\partial r}T_{t}^{t}\left(
r_{+}\right)  +T_{t}^{t}\left(  r_{+}\right)  \left(  3\ell^{2}-r_{+}%
^{2}\left(  d-6\right)  \right)  \right]  .
\end{align}

\subsubsection{With $\sigma=\pm1$ or $\frac{d-1}{2}$ is an even integer
($d=5,9,13,...$)}

It is clear that in this case $\sigma=-1$ does not claim any horizon and
therefore it is out of our interest but for $\sigma=1$ branch the Hawking
temperature and the specific heat capacity are given as (42) and (44) but
there exists an additional constraint on the free parameter in order to have a
real metric function.

To complete this section we give an example for the CSBI black hole $\left(
\chi=1,\sigma=1\right)  $ in $5$-dimensions. To do so we recall that in
$5$-dimensions%
\begin{equation}
B\left(  r\right)  =%
{\textstyle\int^{r}}
z^{3}T_{t}^{t}\left(  z\right)  dz=\frac{\beta^{2}r^{4}}{2}\left(
1-\sqrt{1+\frac{q^{2}}{\beta^{2}r^{6}}}\right)  +\frac{3}{4}\frac{q^{2}}%
{r^{2}}\text{ }_{2}F_{1}\left(  \frac{1}{2},\frac{1}{3},\frac{4}{3}%
,-\frac{q^{2}}{\beta^{2}r^{6}}\right)
\end{equation}
and therefore%
\begin{equation}
f\left(  r\right)  =1+\frac{r^{2}}{\ell^{2}}-\left\{  m+1-\left[  \frac
{2\beta^{2}r^{4}}{3\ell^{2}}\left(  1-\sqrt{1+\frac{q^{2}}{\beta^{2}r^{6}}%
}\right)  +\frac{q^{2}}{\ell^{2}r^{2}}\text{ }_{2}F_{1}\left(  \frac{1}%
{2},\frac{1}{3},\frac{4}{3},-\frac{q^{2}}{\beta^{2}r^{6}}\right)  \right]
\right\}  ^{\frac{1}{2}}.
\end{equation}
In Fig. 1 we plot $f\left(  r\right)  $ for the specific choices of mass,
charge and the cosmological constant. The corresponding temperature and
specific heat capacity plots are given in Fig.s 2 and 3.

\textit{Example 5} \emph{(Clouds of strings as source)}: We consider another
application of the Theorem that incorporates energy-momentum tensor
representing clouds of string type matter fields \cite{12}. More recently
\cite{13}, clouds of string type energy-momentum is considered in
Einstein-Gauss-Bonnet (EGB) gravity. In Rem. 2 and 3, we considered the case
for a general energy - momentum tensor and its corresponding solution. Clouds
of string type matter fields obey the condition imposed on the energy -
momentum tensor stated in Remark 3, such that,%

\begin{align}
T_{t}^{t} &  =T_{r}^{r}=\frac{a}{r^{d-2}},\\
a &  =\text{real constant,}\nonumber
\end{align}
which leads
\begin{equation}
T_{\theta_{i}}^{\theta_{i}}=0.
\end{equation}
Therefore the general solution for the metric function is obtained from the
algebraic equation%
\begin{equation}
\sum_{k=1}^{\left[  \frac{d-1}{2}\right]  }\tilde{\alpha}_{k}H\left(
r\right)  ^{k}=\frac{\Lambda}{3}+\frac{M}{r^{d-1}}-\frac{2a}{\left(
d-2\right)  r^{d-2}}.
\end{equation}
The solution for $H\left(  r\right)  $ generalizes the solution obtained in
Ref.\cite{13} to higher order Lovelock gravity. And hence, our general
solution includes the solution obtained in \cite{13} if we restrict the
spacetime dimension to $d=5$. For this particular case the solution is
obtained from
\begin{equation}
H\left(  r\right)  +\tilde{\alpha}_{2}H\left(  r\right)  ^{2}=\frac{\Lambda
}{3}+\frac{M}{r^{4}}-\frac{2a}{3r^{3}},
\end{equation}
or equivalently%
\begin{equation}
f\left(  r\right)  =\chi+\frac{r^{2}}{4\alpha_{2}}\left(  1\pm\sqrt
{1+8\alpha_{2}\left(  \frac{\Lambda}{3}+\frac{M}{r^{4}}-\frac{2a}{3r^{3}%
}\right)  }\right)  .
\end{equation}
This solution is nothing but the black hole in EGB gravity in the presence of
the string cloud type matter fields \cite{12}. In this theory one can easily
check that the corresponding energy momentum tensor in $d$-dimensions is given
by%
\begin{equation}
T_{\mu}^{\nu}=\text{diag}\left(  \frac{a}{r^{d-2}},\frac{a}{r^{d-2}%
},0,0,...\right)  .
\end{equation}
At this stage we wish to go a step further to relate our solution to the CS
solution in odd $d$-dimensions as ($\chi=1$),%
\begin{equation}
f\left(  r\right)  =1+\frac{r^{2}}{\ell^{2}}-\sigma\left[  m+1-\frac{\left(
d-1\right)  ar}{\left(  d-2\right)  \ell^{\left(  d-3\right)  }}\right]
^{\frac{2}{d-1}},
\end{equation}
which in $5$-dimensions becomes%
\begin{equation}
f\left(  r\right)  =1+\frac{r^{2}}{\ell^{2}}\pm\sqrt{m+1-\frac{4ar}{3\ell^{2}%
}}.
\end{equation}

The thermodynamic properties of this solution is also investigated. The
Hawking temperature ( $T_{H}$) and heat capacity $C_{a}$ at constant string
parameter $a$ is calculated at the location of the event horizon ( $r_{h}$ )
are given by,%

\begin{equation}
T_{H}=\frac{l^{2}(3r_{h}+a)+3r_{h}^{3}}{3\pi l^{2}(l^{2}+r_{h}^{2})},
\end{equation}

\begin{equation}
C_{a}=\frac{3\pi r_{h}^{2}(l^{2}+r_{h}^{2})\left[  l^{2}\left(  3r_{h}%
+a\right)  +3r_{h}^{3}\right]  }{2\left[  l^{2}\left(  6r_{h}^{2}%
+3l^{2}-2ar_{h}\right)  +3r_{h}^{4}\right]  }%
\end{equation}

We also analyzed the thermodynamic stability which is indicated by the
positive heat capacity $C_{a}.$ If the heat capacity has unbounded
discontinuity at particular points of $r_{h}$, this implies possible phase
change from stable to unstable black hole solution. As illustrated in Fig. 4
the transitions from stable to unstable black hole solution is not continuous
and therefore possible Hawking-Page type phase transition occurs \cite{15}.
The occurrence of the phase transition crucially depends on the ratio of
$a/l.$ In Fig. 4, the values of this ratio that creates phase transitions is depicted.

\section{A generalization to the radiating metrics}

Following Bonnor and Vaidya (BV) \cite{8}, we consider a non-linear
electrodynamic Lagrangian $\mathcal{L}\left(  \mathcal{F}\right)  $, a null
current $\mathbf{J}$ and a coupling term $A_{\mu}J^{\mu}$ added to the
original Lagrangian such that
\begin{equation}
d\left(  ^{\star}\mathbf{F}\mathcal{L}_{\mathcal{F}}\right)  =\ ^{\star
}\mathbf{J.}%
\end{equation}
Now, we consider the $d$-dimensional version of the BV metric
\begin{equation}
ds^{2}=-(\chi-r^{2}H\left(  r,u\right)  )du^{2}+2\epsilon drdu+r^{2}%
d\Omega_{d-2}^{2},
\end{equation}
with the outgoing null coordinate $u$ and field 2-form
\begin{equation}
\mathbf{F=}E\left(  u,r\right)  dr\wedge du.
\end{equation}
This gives%
\begin{equation}
^{\star}\mathbf{F=}E\left(  u,r\right)  \sqrt{-g}d\theta_{1}\wedge d\theta
_{2}...\wedge d\theta_{d-2},
\end{equation}
in which $g=\det\left(  g_{\mu\nu}\right)  $ and the Einstein equation is
given by
\begin{equation}
\mathcal{G}_{\mu}^{\nu}=T_{\mu}^{\nu\left(  em\right)  }+T_{\mu}^{\nu\left(
fluid\right)  }.
\end{equation}
Here%
\begin{align}
\mathcal{G}_{\mu}^{\nu} &  =\mathcal{G}_{\mu}^{\ \nu EH}+\sum_{k=2}^{\left[
\frac{d-1}{2}\right]  }\alpha_{k}\mathcal{G}_{\mu}^{\ \nu\left(  k\right)
},\text{ \ \ }\nonumber\\
T_{\mu}^{\nu\left(  em\right)  } &  =\frac{1}{2}\left(  \mathcal{L}\delta
_{\mu}^{\ \nu}-4\mathcal{L}_{\mathcal{F}}F_{\mu\lambda}F^{\nu\lambda}\right)
\end{align}
and
\begin{equation}
T_{\mu}^{\nu\left(  fluid\right)  }=-V^{\nu}V_{\mu}.
\end{equation}
for a null vector $V_{\mu}$. We start now with NLM equation (50) which leads
to%
\begin{equation}%
\begin{tabular}
[c]{l}%
$d\left(  ^{\ast}\mathbf{F}\mathcal{L}_{\mathcal{F}}\right)  =d\left(
E\left(  u,r\right)  \mathcal{L}_{\mathcal{F}}\sqrt{-g}d\theta_{1}\wedge
d\theta_{2}...\wedge d\theta_{d-2}\right)  =$\\
$\left[  \left(  E\left(  u,r\right)  \mathcal{L}_{\mathcal{F}}\sqrt
{-g}\right)  _{r}dr+\left(  E\left(  u,r\right)  \mathcal{L}_{\mathcal{F}%
}\sqrt{-g}\right)  _{u}du\right]  \wedge d\theta_{1}\wedge d\theta
_{2}...\wedge d\theta_{d-2}=^{\ast}\mathbf{J.}$%
\end{tabular}
\end{equation}
By using the relation between $1$-form current $\mathbf{J}$ and its dual i.e.,
$\mathbf{J=}\left(  \mathbf{-1}\right)  ^{d\ \ \ast\ast}\mathbf{J}$ one finds%
\begin{align}
\mathbf{J} &  \mathbf{=}\left(  \mathbf{-1}\right)  ^{d}\left[  \left(
E\left(  u,r\right)  \mathcal{L}_{\mathcal{F}}\sqrt{-g}\right)  _{r}\ ^{\star
}\left(  dr\wedge d\theta_{1}\wedge d\theta_{2}...\wedge d\theta_{d-2}\right)
\right.  +\nonumber\\
&  \left.  \left(  E\left(  u,r\right)  \mathcal{L}_{\mathcal{F}}\sqrt
{-g}\right)  _{u}\ ^{\star}\left(  du\wedge d\theta_{1}\wedge d\theta
_{2}...\wedge d\theta_{d-2}\right)  \right]  .
\end{align}
From the metric we find%
\begin{gather}
^{\star}\left(  dr\wedge d\theta_{1}\wedge d\theta_{2}...\wedge d\theta
_{d-2}\right)  =\frac{\left(  -1\right)  ^{d-1}}{\sqrt{-g}}\left(  dr-\epsilon
fdu\right)  ,\text{ \ \ \ \ \ \ \ \ }\\
\text{ }^{\star}\left(  du\wedge d\theta_{1}\wedge d\theta_{2}...\wedge
d\theta_{d-2}\right)  =\frac{\left(  -1\right)  ^{d-1}}{\sqrt{-g}}du,
\end{gather}
and therefore%
\begin{equation}
\mathbf{J}\mathbf{=}\left(  E\left(  u,r\right)  \mathcal{L}_{\mathcal{F}%
}r^{d-2}\right)  _{r}\ \frac{1}{r^{d-2}}\left(  dr-\epsilon fdu\right)
-\left(  E\left(  u,r\right)  \mathcal{L}_{\mathcal{F}}\right)  _{u}\ du.
\end{equation}
This current is going to be null i.e., $J_{\mu}=$ $\left(  J_{u}%
,0,...,0\right)  =J_{u}\delta_{\mu}^{u}$ which means that $\left(  E\left(
u,r\right)  \mathcal{L}_{\mathcal{F}}r^{d-2}\right)  _{r}=0$ integrates to%
\begin{equation}
E\left(  u,r\right)  \mathcal{L}_{\mathcal{F}}r^{d-2}=Q\left(  u\right)
\end{equation}
or%
\begin{equation}
E\left(  u,r\right)  \mathcal{L}_{\mathcal{F}}=\frac{Q\left(  u\right)
}{r^{d-2}},
\end{equation}
for a $u$ dependent charge $Q(u)$. After considering these results we find
\begin{align}
\mathbf{F} &  =\frac{Q\left(  u\right)  }{r^{d-2}\mathcal{L}_{\mathcal{F}}%
}dr\wedge du,\\
\mathbf{J} &  \mathbf{=}\mathbf{-}\left(  -1\right)  ^{d-1}\left(  E\left(
u,r\right)  \mathcal{L}_{\mathcal{F}}\right)  _{u}du=\left(  -1\right)
^{d}\frac{\dot{Q}\left(  u\right)  }{r^{d-2}}\ du.
\end{align}
where $\dot{Q}\left(  u\right)  =\frac{dQ\left(  u\right)  }{du}.$ The
explicit form of the Maxwell field can be expressed by
\begin{equation}
F\mathcal{=}F_{\mu\nu}F^{\mu\nu}=2F_{ru}F^{ru}=2F_{ru}\left(  g^{\alpha
r}g^{\beta u}F_{\alpha\beta}\right)  .
\end{equation}
while our metric tensor $g_{\mu\nu}$ and $g^{\mu\nu}$ are
\begin{equation}%
\begin{tabular}
[c]{l}%
$g_{\mu\nu}=\left(
\begin{array}
[c]{cccccc}%
-f & \epsilon & 0 & 0 & . & .\\
\epsilon & 0 & 0 & 0 & . & .\\
0 & 0 & r^{2} & 0 & . & .\\
0 & 0 & 0 & r^{2}\sin^{2}\theta & . & .\\
. & . & . & . & . & .\\
. & . & . & . & . & .
\end{array}
\right)  ,$%
\end{tabular}
\ \ \ \ \
\begin{tabular}
[c]{l}%
$g^{\mu\nu}=\left(
\begin{array}
[c]{cccccc}%
0 & \frac{1}{\epsilon} & 0 & 0 & . & .\\
\frac{1}{\epsilon} & f & 0 & 0 & . & .\\
0 & 0 & \frac{1}{r^{2}} & 0 & . & .\\
0 & 0 & 0 & \frac{1}{r^{2}\sin^{2}\theta} & . & .\\
. & . & . & . & . & .\\
. & . & . & . & . & .
\end{array}
\right)  $%
\end{tabular}
\end{equation}
giving
\begin{equation}
\mathcal{F=}-2\left(  F_{ru}\right)  ^{2}.
\end{equation}
The energy - momentum tensor components are given by%
\begin{align}
&
\begin{tabular}
[c]{l}%
$T_{u}^{u\left(  em\right)  }=\frac{1}{2}\left(  \mathcal{L}-4\mathcal{L}%
_{\mathcal{F}}F_{u\lambda}F^{u\lambda}\right)  =\frac{1}{2}\left(
\mathcal{L}-2\mathcal{L}_{\mathcal{F}}\mathcal{F}\right)  ,$%
\end{tabular}
\\
&
\begin{tabular}
[c]{l}%
$T_{r}^{r\left(  em\right)  }=\frac{1}{2}\left(  \mathcal{L}-4\mathcal{L}%
_{\mathcal{F}}F_{r\lambda}F^{r\lambda}\right)  =\frac{1}{2}\left(
\mathcal{L}-2\mathcal{L}_{\mathcal{F}}\mathcal{F}\right)  =T_{u}^{u\left(
em\right)  },$%
\end{tabular}
\\
&
\begin{tabular}
[c]{l}%
$T_{\theta_{i}}^{\theta_{i}\left(  em\right)  }=\frac{1}{2}\mathcal{L},$%
\end{tabular}
\\
&  T_{u}^{r\left(  em\right)  }=T_{r}^{u\left(  em\right)  }=0.
\end{align}
The null-fluid current vector is $V_{\mu}\mathbf{=}\left(  V_{u}%
,0,...,0\right)  =V_{u}\delta_{\mu}^{u}$ and therefore
\begin{equation}
V^{\mu}=\epsilon\delta_{r}^{\mu}V_{u},
\end{equation}
implying that
\begin{equation}
g_{\mu\nu}V^{\mu}V^{\nu}=g_{rr}\left(  V^{r}\right)  ^{2},
\end{equation}
which obviously vanishes and
\begin{equation}
T_{\mu}^{\nu\left(  fluid\right)  }=-V^{\nu}V_{\mu}=-V^{r}V_{u}\delta_{r}%
^{\nu}\delta_{\mu}^{u}=-\epsilon\left(  V_{u}\right)  ^{2}\delta_{r}^{\nu
}\delta_{\mu}^{u}.
\end{equation}
Finally, we give the explicit form of the energy momentum tensor as%
\begin{equation}
T_{\ \mu}^{\nu}=\left(
\begin{array}
[c]{cccccc}%
\frac{1}{2}\left(  \mathcal{L}-2\mathcal{L}_{\mathcal{F}}\mathcal{F}\right)
& 0 & 0 & 0 & . & .\\
-\epsilon\left(  V_{u}\right)  ^{2} & \frac{1}{2}\left(  \mathcal{L}%
-2\mathcal{L}_{\mathcal{F}}\mathcal{F}\right)   & 0 & 0 & . & .\\
0 & 0 & \frac{1}{2}\mathcal{L} & 0 & . & .\\
0 & 0 & 0 & \frac{1}{2}\mathcal{L} & . & .\\
. & . & . & . & . & .\\
. & . & . & . & . & .
\end{array}
\right)  .
\end{equation}
Similarly the Einstein tensor can be expressed by%
\begin{equation}
G_{\ \mu}^{\nu}=\left(
\begin{array}
[c]{ccccc}%
-\frac{d-2}{2r^{d-2}}\left(  r^{d-1}%
{\textstyle\sum\limits_{i=1}^{\left[  \frac{d-1}{2}\right]  }}
\tilde{\alpha}_{i}H^{i}\right)  ^{\prime} & 0 & 0 & . & .\\
\frac{d-2}{2}r\frac{\partial}{\partial u}%
{\textstyle\sum\limits_{i=1}^{\left[  \frac{d-1}{2}\right]  }}
\tilde{\alpha}_{i}H^{i} & -\frac{d-2}{2r^{d-2}}\left(  r^{d-1}%
{\textstyle\sum\limits_{i=1}^{\left[  \frac{d-1}{2}\right]  }}
\tilde{\alpha}_{i}H^{i}\right)  ^{\prime} & 0 & . & .\\
0 & 0 & -\frac{1}{2r^{d-3}}\left(  r^{d-1}%
{\textstyle\sum\limits_{i=1}^{\left[  \frac{d-1}{2}\right]  }}
\tilde{\alpha}_{i}H^{i}\right)  ^{\prime\prime} & . & .\\
0 & 0 & 0 & . & .\\
. & . & . & . & .\\
. & . & . & . & .
\end{array}
\right)
\end{equation}
where%
\begin{equation}
f\left(  u,r\right)  =1-r^{2}H\left(  u,r\right)  .
\end{equation}
After all these arrangements we are ready now to solve the Einstein equations.
We start with the $uu$ component
\begin{equation}
-\frac{d-2}{2r^{d-2}}\left(  r^{d-1}%
{\textstyle\sum\limits_{i=1}^{\left[  \frac{d-1}{2}\right]  }}
\tilde{\alpha}_{i}H^{i}\right)  ^{\prime}=-\left(  \mathcal{L}-2\mathcal{L}%
_{\mathcal{F}}\mathcal{F}\right)  \rightarrow%
{\textstyle\sum\limits_{i=1}^{\left[  \frac{d-1}{2}\right]  }}
\tilde{\alpha}_{i}H^{i}=\frac{M(u)}{r^{d-1}}-\frac{2}{\left(  d-2\right)
r^{d-1}}\int r^{d-2}T_{r}^{r}dr.
\end{equation}
One can easily check that the $rr$ component gives the same result while the
angular components give%
\begin{equation}
-\frac{1}{2r^{d-3}}\left(  r^{d-1}%
{\textstyle\sum\limits_{i=1}^{\left[  \frac{d-1}{2}\right]  }}
\tilde{\alpha}_{i}H^{i}\right)  ^{\prime\prime}=\frac{1}{2}%
\mathcal{L\rightarrow}\left(  \frac{r^{d-2}}{d-2}\left(  \mathcal{L}%
-2\mathcal{L}_{\mathcal{F}}\mathcal{F}\right)  \right)  ^{\prime}%
=r^{d-3}\mathcal{L}%
\end{equation}
which is nothing but the NLM equation (14), already satisfied. In the last
step we work on $\mathcal{G}_{r}^{u}=T_{r}^{u}$ giving%
\begin{equation}
\frac{d-2}{2}r\frac{\partial}{\partial u}%
{\textstyle\sum\limits_{i=1}^{\left[  \frac{d-1}{2}\right]  }}
\tilde{\alpha}_{i}H^{i}=-\epsilon\left(  V_{u}\right)  ^{2}%
\end{equation}
and therefore%
\begin{equation}
\frac{d-2}{2}r\left(  \frac{\dot{M}(u)}{r^{d-1}}-\frac{2}{\left(  d-2\right)
r^{d-1}}\int r^{d-2}\dot{T}_{r}^{r}dr\right)  =-\epsilon\left(  V_{u}\right)
^{2}%
\end{equation}
which finally determines the null-fluid current vector by%
\begin{equation}
V_{u}^{2}=-\epsilon\left(  \frac{\dot{M}(u)}{r^{d-1}}-\frac{1}{\left(
d-2\right)  r^{d-1}}\partial_{u}%
{\textstyle\int}
drr^{d-2}\left(  \mathcal{L}-2\mathcal{L}_{\mathcal{F}}\mathcal{F}\right)
\right)  .
\end{equation}

\section{Radiating power-Maxwell source in Lovelock gravity}

In this section our choice for the NLM field consists of a particular kind,
namely%
\begin{equation}
\mathcal{L}\left(  \mathcal{F}\right)  =-\mathcal{F}^{q}%
\end{equation}
in which $q$ stands for a constant parameter and $\mathcal{F=}F_{\mu\nu}%
F^{\mu\nu}$, as before. Let us note that this particular version of
non-linearity attracted considerable interest in recent years \cite{14}.
Although the parameter $q$ (i.e. the power) may be assumed arbitrary,
imposition of the energy - conditions with other requirements restrict $q$ to
a limited set of values. The rest of the gravitational action will be chosen
as in the previous sections. Similarly, our choice of the line element follows
that of (60). The electromagnetic field 2-form turns out to be%

\begin{equation}
F=\frac{Q(u)}{r^{d-2}\mathcal{L}_{\mathcal{F}}}dr\wedge du,
\end{equation}
and energy-momentum component $T_{r}^{\text{ \ }r}$
\begin{equation}
T_{r}^{\text{ \ }r}=\frac{1}{2}\left(  \mathcal{L}-2\mathcal{L}_{\mathcal{F}%
}\mathcal{F}\right)  =\left(  q-\frac{1}{2}\right)  \mathcal{F}^{q}\text{ ,
\ \ }\text{\ \ }\left(  q\neq\frac{1}{2}\right)  ,
\end{equation}
so that%

\begin{equation}
\mathcal{F}^{q}=\left(  -1\right)  ^{q}\left(  \frac{2Q^{2}(u)}{q^{2}%
r^{2\left(  d-2\right)  }}\right)  ^{\frac{q}{2q-1}}.
\end{equation}

Following the procedure of the previous section further, which led us to the
condition (88) we obtain now, in analogy%

\begin{equation}%
{\textstyle\sum\limits_{i=1}^{\left[  \frac{d-1}{2}\right]  }}
\tilde{\alpha}_{i}H^{i}=\frac{M(u)}{r^{d-1}}-h_{d}(u)r^{\frac{2\left(
q-d+1\right)  }{2q-1}}%
\end{equation}
in which%

\begin{equation}
h_{d}(u)=\left(  -1\right)  ^{q}\left(  \frac{2q-1}{d-2}\right)  \left(
\frac{2Q^{2}(u)}{q^{2}}\right)  ^{\frac{q}{2q-1}}.
\end{equation}
It is observed that an arbitrary $q$ does not guarantee the reality of the
metric function, which enforces us to choose $q$ appropriately.

The particular dimensionality $d=5$ brings in significant simplicity to the
foregoing expressions such as%

\begin{equation}
\tilde{\alpha}_{1}H+\tilde{\alpha}_{2}H^{2}=\frac{M(u)}{r^{4}}-h_{5}%
(u)r^{\frac{2\left(  q-4\right)  }{2q-1}},
\end{equation}
where $\tilde{\alpha}_{1}=1$ and $\tilde{\alpha}_{2}=2\alpha$. The quadratic
equation for $H$ can easily be solved and the corresponding metric function
$f(r,u)$ is determined by%

\begin{equation}
f(r,u)=\chi+\frac{r^{2}}{4\alpha}\left(  1\pm\sqrt{1+8\alpha\left(
\frac{M(u)}{r^{4}}-h_{5}(u)r^{\frac{2\left(  q-4\right)  }{2q-1}}\right)
}\right)  .
\end{equation}

In analogy to Eq. (82), the null-current component for the present case turns
out to be%

\begin{equation}
V_{u}^{2}=-\epsilon\frac{3}{2}r\left(  \frac{\dot{M}(u)}{r^{4}}-\dot{h}%
_{5}(u)r^{\frac{2\left(  q-4\right)  }{2q-1}}\right)
\end{equation}
in which, as usual, a 'dot' represents $\frac{d}{du}.$

To be able to proceed further with the thermodynamical properties in the
present choice of $d=5$ $($ and $\chi=+1)$, we must determine the apparent
horizon through $f(r,u)=0.$ This leads us to the algebraic relation%

\begin{equation}
2\alpha-M(u)+r_{h}^{2}+h_{5}(u)r_{h}^{k}=0
\end{equation}
in which $r_{h}$ denotes apparent horizon (if any) and we have abbreviated%

\begin{equation}
k=2\left(  \frac{5q-6}{2q-1}\right)  .
\end{equation}

Tab. 1 shows the relation between $q$ and $k$ for certain leading numbers of
interest. With reference to Table 1, we can find a finely-tuned set of $q$
powers so that $h_{5}(u)$ will be real. Clearly, any odd/even integer $q$ will
do the job whereas non-integer $q$'s will not serve the purpose. For instance,
$q=4$ is a good choice. Accordingly, we find%

\begin{equation}
h_{5}(u)=\frac{7}{3}\left(  \frac{Q^{2}\left(  u\right)  }{8}\right)
^{4/7}>0,
\end{equation}

and%

\begin{align}
r_{h}\left(  u\right)   &  =\frac{1}{2}\left(  \sqrt{1+4h_{5}(u)\left(
M(u)-2\alpha\right)  }-1\right) \\
M(u)  &  >2\alpha.\nonumber
\end{align}

The corresponding metric function takes the form%

\begin{equation}
f(r,u)=1+\frac{r^{2}}{4\alpha}\left(  1\pm\sqrt{1+8\alpha\frac{M(u)}{r^{4}%
}-h_{5}(u)}\right)  .
\end{equation}

Different choices of $q$ values from Table 1 can be treated in a similar
manner to obtain the corresponding $f(r,u)$ function, which we shall not go
any further in this paper.

\section{Conclusion}

Methods to generate new solutions for various energy-momenta are available in
the literature. In this regard, recently we have proved a theorem that
generalizes Salgado's theorem \cite{7}. The whole issue in such a problem is
the physical significance of the energy-momentum under consideration. Scalar
field source, for instance, is known to work only in $d=4$. Linear Maxwell
electrodynamics has already been incorporated in higher dimensional Lovelock
gravity \cite{2,4}. Our example 5 which is a particular version of the Theorem
concerns a cloud of strings without Maxwell field. We explore thermodynamic
properties of such a cloud for a particular solution in $d=5$ spacetime
dimension, representing Chern-Simon black holes which undergoes a Hawking-Page
phase transition (Fig. 4). In this paper we extended the problem to cover a
non-linear Maxwell (NLM) source. Among this class of theories Born-Infeld (BI)
is the most familiar one, which arises as a particular example to the Theorem.
The Chern-Simons-Born-Infeld (CSBI) metrics that we found by using our Theorem
turn out to be thermodynamically stable. Next, we address to dynamical problem
that covers time dependent metrics and NLM fields. Stated otherwise, we have
generalized the well-known dynamical Bonnor-Vaidya (BV) metric to the general
Lovelock theory with NLM sources. A radiating null current source naturally
accompanies the radiating energy - momentum of such metrics which lose mass
and charge. It is needless to remark, finally, that the opposite problem of
the 'shining star', namely, the collapse of time dependent energy - momentum
and null radiation current is also solved in the same theory.

\bigskip

\textbf{Captions:}

Table 1: The relation between the parameters $k$ and $q$ according to Eq. (103).

Figure 1: The plot of $f\left(  r\right)  $ (Eq. (48)), for the specific
parameters $m=\ell=q=1$, as $\beta$ ranges from $0$ to $\infty$. It is seen
that black holes solutions are available only for $\beta\leq\beta
_{critical}=0.2276.$

Figure 2: Hawking temperature $T_{H}$ versus event horizon radius $r_{+}$ for
the specific parameters $\ell=q=1$ and for $\beta\in\left(  0,\beta
_{critical}\right]  .$ By taking the absolute value of $T_{H}$ automatically
deletes the negative temperatures as non-physical.

Figure 3: specific heat capacity $C_{q}$ versus $r_{+}$ for versus $r_{+}$ for
$\ell=q=1$ and $\beta\in\left(  0,\beta_{critical}\right]  .$ This plot,
(together with Fig. 1), reveals that our CSBI black hole solution is
thermodynamically stable.

Figure 4: Heat Capacity $C_{a}$ versus the horizon radius $r_{h}$ for specific
parameters $a=5$ and $\ell=1.$ The singularity in $C_{a}$ and therefore
occurrence of Hawking-Page phase transition is clearly seen. The dash region
in the inscribed figure depicts $\frac{a}{\ell}$ versus $r_{h}$ for which such
a transition occurs.

\textbf{Table 1:}
\[%
\begin{tabular}
[c]{|c|c|c|c|c|c|c|c|c|c|}\hline
$k$ & 0 & 1 & 2 & 3 & 4 & 6 & . & . & .\\\hline
$q$ & $\frac{6}{5}$ & $\frac{11}{8}$ & $\frac{5}{3}$ & $\frac{9}{4}$ & 4 &
-3 & . & . & .\\\hline
\end{tabular}
\ \
\]

\end{document}